\documentclass[aps,pra,letterpaper,notitlepage,longbibliography,floatfix,12pt,reprint]{revtex4-1}

\usepackage{CJK} 

\usepackage{amsmath,amsfonts,amssymb}

\usepackage{float} 

\usepackage{graphicx}
\usepackage[subrefformat=parens, caption=false]{subfig}

\usepackage{hyperref}
\hypersetup{colorlinks=true}
\usepackage[all]{hypcap} 

\usepackage{times}

\usepackage{upgreek}

\usepackage{color}

\newcommand{\figurepanel}[2]{\hyperref[#1]{\ref*{#1}(#2)}}

\allowdisplaybreaks[1] 

\usepackage{filecontents}

\begin{document}
\begin{CJK*}{UTF8}{} 
\title{Multiple Emitters in a Waveguide: \\ Non-Reciprocity and Correlated Photons at Perfect Elastic Transmission}
\CJKfamily{bsmi}
\author{Yao-Lung L. Fang (方耀龍)}
\author{Harold U. Baranger}
\affiliation{Department of Physics, Duke University, P.O. Box 90305, Durham, North Carolina 27708-0305, USA}
\date{June 26, 2017}  

\begin{abstract}
We investigate interference and correlation effects when several detuned emitters are placed along a one-dimensional photonic waveguide. Such a setup allows multiple interactions between the photons and the strongly coupled emitters, and underlies proposed devices for quantum information processing. We show, first, that a pair of 
detuned two-level systems (2LS) separated by a half wavelength 
mimic a driven $\Lambda$-type three-level system (3LS) in both the single- and two- photon sectors.
There is an interference-induced transparency peak at which
the fluorescence is quenched, leaving the transmitted photons completely uncorrelated. Slightly away from this separation, we find that the inelastic scattering (fluorescence) is large, leading to nonlinear effects such as non-reciprocity (rectification). We connect this non-reciprocity to inelastic scattering caused by driving a dark pole and so derive a condition for maximum rectification. 
Finally, by placing a true 3LS midway between the two 2LS, we show that elastic scattering produces only transmission, but inelastic scattering nevertheless occurs (the fluorescence is not quenched) causing substantial photon correlations. 
\end{abstract}

\maketitle
\end{CJK*} 

\section{Introduction}
Stimulated by the strong light-matter interaction that can now be achieved between photons confined in a one-dimensional (1D) channel and local emitters (qubits), the study of waveguide quantum electrodynamics (QED) has received considerable attention \cite{LodahlRMP15,RoyRMP17,NohRPP16,LiaoPhyScr16,WhalenGrimsmoX17,CombesSLH_X16}.
Likewise, driven by the needs of quantum information and computation to transmit and process quantum information using photons, it is of great interest to generate, store, and release single photons in an integrated photonic circuit. Waveguide QED is a natural way to approach these needs and is being actively pursued in a variety of platforms. One recent development, for instance, is the study of directional coupling between the photons and emitters in optical systems \cite{SayrinPRX15,SollnerNNano15,Gonzalez-BallesteroPRA16,LodahlReviewNature17}. Another is the rapid experimental progress in superconducting circuits  \cite{AstafievSci10,AbdumalikovPRL10,EichlerPRL11,HoiPRL12,vanLooScience13,HoiNatPhy15} based on which fundamental quantum system building blocks have been proposed, such as single-photon generators \cite{HouckNat07}, routers \cite{HoiPRL11}, detectors and counters \cite{daSilvaPRA10,ChenPRL11,SathyamoorthyPRL14,InomataNatComm16}, diodes \cite{RoyPRB10,FratiniPRL14}, 
memory \cite{ShenPRB07b,LeungPRL12}, and gates \cite{CiccarelloPRA12,ZhengPRL13a,PaulischNJP16}. 

An important advantage of the 1D waveguide geometry is the ability to attach several local emitters to the waveguide such that each photon  strongly interacts with all of the emitters. Indeed, this feature is behind many of the building blocks mentioned above, and  
in addition paves the way for investigating many-body physics \cite{LalumierePRA13,LaaksoPRL14,PletyukhovPRA15,RoyRMP17,NohRPP16,LiaoPhyScr16} and generating entanglement among the qubits \cite{TudelaPRL11,ZhengPRL13,RochPRL14,GonzalezBallesteroPRA14,PichlerPRL16}. The simplest such 
system consists of two two-level systems (2LS) attached to the waveguide. 
Recent work has shown that when the two 2LS are detuned from each other, rectification of incoming photonic pulses is possible \cite{FratiniPRL14,DaiPRA15,FratiniPRA16,MascarenhasPRA16}.

In this paper, we present two dramatic effects of 
detuned emitters in waveguide QED. First, we show that two 2LS coupled to the waveguide can act like a driven $\Lambda$-type three-level system (3LS), a highly desirable structure, not only for a single photon but also in the \emph{two-photon} sector. At certain qubit separations $L$, full transmission of photons within a narrow frequency range is achieved via interference, as in electromagnetically induced transparency (EIT) in a $\Lambda$-3LS \cite{FleischhauerRMP05}. Similar to such a 3LS \cite{ZhengPRA12,RoyPRA14,FangPE16}, at this transmission peak, the photons are not correlated. Indeed, there is no inelastic scattering (the fluorescence is quenched) and no bunching or anti-bunching.  

Strong inelastic scattering can however be produced by slightly changing the qubit separation around the ``EIT point.'' This occurs because the dark state associated with EIT acquires a small decay width, producing inelastic scattering. We show that this is intimately connected to the non-reciprocity  discussed in Refs.~\cite{DaiPRA15,FratiniPRA16,MascarenhasPRA16}. In particular, the linear relation between parameters that maximize rectification reported in Ref.~\cite{DaiPRA15} can be explained assuming the dark state is driven. Our finding provides useful guidance in designing photonic rectifiers.

\begin{figure}[t]
	\centering
	\includegraphics[scale=0.9, trim = 0 1.7cm 0 2cm, clip=true]{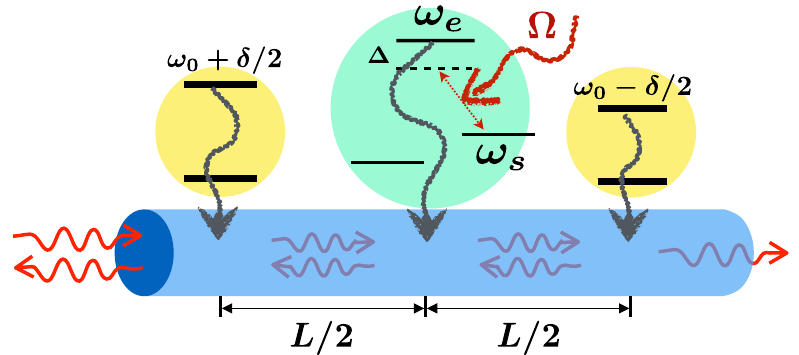}
	\caption{Schematic for the ``2-3-2'' structure: a pair of  2LS, separated by distance $L$ and detuned by $\delta$, coupled to a 1D waveguide, with an additional driven $\Lambda$-type 3LS placed in the middle. In the rotating frame, $\omega_s=\omega_e-\Delta$ where $\Delta$ is the detuning of the classical beam. Throughout this paper, we assume that two identical photons with momentum $k$ are incident.
		\label{fig:schematic}}
\end{figure}

The second multiple-emitter effect presented here is that the behavior is strikingly different when a genuine driven $\Lambda$-3LS is inserted between the two 2LS, hereafter referred to as the ``2-3-2'' structure shown in Fig.\,\ref{fig:schematic}. In the combined structure, there continues to be perfect transmission in the elastic channel, but now inelastic scattering produces photons that are strongly correlated. The resulting resonance fluorescence and bunching in the photon-photon correlations, $g_2(0) \!>\! 1$, are both large. 

All of our results are obtained using a scattering approach developed previously \cite{ZhengPRL13,FangEPJQT14,FangPRA15}, which is equivalent to the weak-pumping limit in input-output theory \cite{FangPRA15}.  An important tool is the use of the total inelastic photon flux as a figure of merit to find situations and parameters where the photon correlations are strongest \cite{FangPE16}. 

The rest of this paper is organized as follows: In Sec.~\ref{sec: the model} we describe the Hamiltonian of the systems under consideration. Next, we discuss in Sec.~\ref{sec: mapping} the mapping between a pair of detuned 2LS and a driven $\Lambda$-3LS under the Markovian approximation, and then move on to results in first- and second-order quantities such as single-photon transmittance $T(k)$, two-photon inelastic-scattering flux $F(k)$, two-photon correlation function $g_2(t)$, and time delay $\uptau$. 
The recent proposal on rectification effects \cite{FratiniPRL14,DaiPRA15,FratiniPRA16,MascarenhasPRA16} is revisited in Sec.~\ref{sec: rectification}. In Sec.~\ref{sec: 2-3-2}, we discuss how to suppress the fluorescence quench by interrupting the 2LS pair with a genuine driven $\Lambda$-3LS, and show that the transmitted photons have nontrivial correlations. We close in Sec.~\ref{sec:loss-markovian} with a brief discussion of the validity of Markovian approximation and the robustness of this mapping with nonzero dissipation.

\section{The model}
\label{sec: the model}

We consider a 1D continuum with linear dispersion and bi-directional photons denoted by ``R'' and ``L'' for right-moving and left-moving, respectively. The Hamiltonian in position space is, then, 
\begin{equation}
\mathcal{H}_\text{ph}=-i \!\int \!\!dx
\left[ a_R^\dagger(x) \frac{d}{dx} a_R(x) - a_L^\dagger(x) \frac{d}{dx} a_L(x) \right].
\end{equation}
(For convenience, we take $\hbar\!=\!c\!=\!1$.) 
Each of the two detuned 2LS is characterized by a transition frequency $\omega_i$, decay rate $\Gamma_i$, position $x_i$, and raising/lowering operator $\sigma_{i\pm}$. The Hamiltonian of the 2LS is, then, 
$\mathcal{H}^0_\text{2LS} = \sum_{i}\omega_i \sigma_{i+}\sigma_{i-}$.
We take (without loss of generality) $\omega_{1,2}=\omega_0\pm\delta/2$ and $x_{1,2}=\pm L/2$.  

When a pair of distant 2LS is coupled to the waveguide, the full Hamiltonian is 
$\mathcal{H}_\text{2-2LS}=\mathcal{H}_\text{ph}+\mathcal{H}^0_\text{2LS}+\mathcal{H}_\text{2-2LS}^\text{int}$, where  
\begin{equation}
   \mathcal{H}_\text{2-2LS}^\text{int} = 
   \sum_{\substack{\alpha=\text{R,L}\\i=1,2}} 
   \sqrt{\frac{\Gamma_i}{2}} \int dx\, \delta(x-x_i) 
   \left[a_\alpha^\dagger(x)\sigma_{i-}+ \text{h.c.}\right]
\end{equation}
in the rotating wave approximation.

For the 2-3-2 structure, a single driven $\Lambda$-type 3LS is added at $x=0$ (the middle element in Fig.\,\ref{fig:schematic}). Its Hamiltonian is 
$\mathcal{H}^0_\text{3LS} = 
\sum_{\beta=e,s} (\omega_\beta \sigma_{\beta+}\sigma_{\beta-})
+ \Omega/2 (\sigma_{e+}\sigma_{s-} + \text{h.c.})$, 
where ``e'' and ``s'' refer to the excited and metastable states of the 3LS, 
$\omega_s=\omega_e-\Delta$ in the rotating frame, and $\Omega$ ($\Delta$) is the Rabi frequency (detuning) of the classical beam driving the \textit{e-s} transition.  
The coupling of the 3LS to the waveguide is given by
\begin{equation}
\mathcal{H}_\text{3LS}^\text{int}= \sum_{\alpha=\text{R,L}}\sqrt{\frac{\Gamma_e}{2}} \int dx\, \delta(x) \left[a_\alpha^\dagger(x)\sigma_{e-}+\text{h.c.}\right].
\end{equation}
Thus, the full Hamiltonian of the 2-3-2 structure is given by
$\mathcal{H}_\text{2-3-2} = \mathcal{H}_\text{ph}+
\mathcal{H}^0_\text{2LS}+ \mathcal{H}^0_\text{3LS} 
+\mathcal{H}_\text{2-2LS}^\text{int}+\mathcal{H}_\text{3LS}^\text{int}$.
\vspace*{0.05in}

\section{Two 2LS vs.\ a $\Lambda$-3LS: Similarities and differences} 
\label{sec: mapping} 

In this section we compare two 2LS with a single driven $\Lambda$-3LS in both the single and two photon sectors. We present the transmission, inelastic scattering, and photon-photon correlation. When the separation between the two 2LS is a multiple of $\lambda_0/2$, there is a clear mapping in the single-photon sector between the two systems. Here the detuning of the two 2LS plays the role of the classical driving field in the $\Lambda$-3LS. 

The main result of this section is that, surprisingly, this mapping carries over to the two-photon sector as well. For resonant photons in particular, interference produces a peak of perfect transmission at which there is no inelastic scattering nor photon correlation. 

\subsection{Single-photon sector}

Using standard methods, the single-photon transmission amplitude $t(k)$ can be obtained for all these systems. For a pair of 2LS, it is (see, for example, \cite{ShenPRB07b,ZhengPRL13})
\begin{equation}
t(k)=\frac{(k-\omega_1) (k-\omega_2)}{(k-\omega_1+i \frac{\Gamma_1}{2}) (k-\omega_2+i \frac{\Gamma_2}{2}) +\frac{\Gamma_1 \Gamma_2}{4} e^{2 i k L}},
\label{eq:transmission amplitude of a pair of 2LS}
\end{equation}
while for a driven $\Lambda$-type 3LS, it is \cite{WitthautNJP10,ZhengPRA12}
\begin{equation}
t(k)=\frac{\left(k-\omega _s\right) \left(k-\omega _e\right)-\frac{\Omega ^2}{4}}{\left(k-\omega _s\right) \left(k-\omega _e+\frac{i \Gamma_e }{2}\right)-\frac{\Omega ^2}{4}}.
\label{eq:transmission amplitude of driven 3LS}
\end{equation}
Unless otherwise stated, we assume the incoming photon with momentum $k$ is injected from the left, as depicted in Fig.~\ref{fig:schematic}. With regard to non-reciprocity, because Eq.~\eqref{eq:transmission amplitude of a pair of 2LS} is invariant upon exchanging subscripts 1 and 2, 
it is evident that a pair of 2LS cannot rectify single photons for any separation $L$ \cite{DaiPRA15,FratiniPRA16,MascarenhasPRA16}. 

We now apply the Markovian approximation to Eq.~\eqref{eq:transmission amplitude of a pair of 2LS} by replacing the propagation factor $\exp(2ikL)$ by $\exp(2ik_0L)$ where $k_0$ is the average wavevector $(\omega_1+\omega_2)/2$. This implementation of the Markovian approximation in a scattering theory approach \cite{ZhengPRL13,FangEPJQT14} is equivalent to those made in master equation approaches. Its validity is ensured when $\Gamma L/c \ll 1$ and $|k-k_0| \alt \Gamma$ as then the phase difference is very small. The Markovian approximation is discussed further in Sec.\,\ref{sec:loss-markovian}. 

We introduce a set of ``mapping rules'' to clarify the relation between transmission through two 2LS and that through a driven $\Lambda$-3LS:
\begin{align}
\omega_e= \omega_0+ \frac{\delta  (\Gamma_1-\Gamma_2)}{2 (\Gamma_1+\Gamma_2)},
&\quad \omega_s= \omega_0-\frac{\delta  (\Gamma_1-\Gamma_2)}{2 (\Gamma_1+\Gamma_2)},\nonumber\\
\Omega = \frac{2 \delta  \sqrt{\Gamma_1 \Gamma_2}}{\Gamma_1+\Gamma_2},
&\quad \Gamma_e = \Gamma_1+\Gamma_2.
\label{eq:mapping rules}
\end{align}
Using these rules while making the particular choice $k_0L=n\pi$ with $n$ an integer, we find that the two amplitudes (\ref{eq:transmission amplitude of a pair of 2LS}) and (\ref{eq:transmission amplitude of driven 3LS}) are \emph{identical}. Notice that the colocated case $n=0$ is included and that the separation between the two 2LS is a multiple of half wavelengths,  $L=n\lambda_0/2$. For simplicity in interpreting this result, we assume hereafter that the two 2LS have the same decay rate, $\Gamma_{1,2}=\Gamma$. When mimicking a driven $\Lambda$-3LS using a pair of 2LS, then, the frequency of the ``excited state'' is the average frequency $\omega_0=(\omega_1+\omega_2)/2$, and the ``Rabi frequency'' $\Omega$ is controlled by the 2LS detuning $\delta$. Ref.~\cite{ShenPRB07b} has a similar discussion in the $L\ll\lambda_0$ limit;
see also Ref.~\cite{LiaoPRA16}.

\begin{figure}[t]
	\centering
	\includegraphics[scale=0.75, trim=0 1cm 0 0, clip=true]{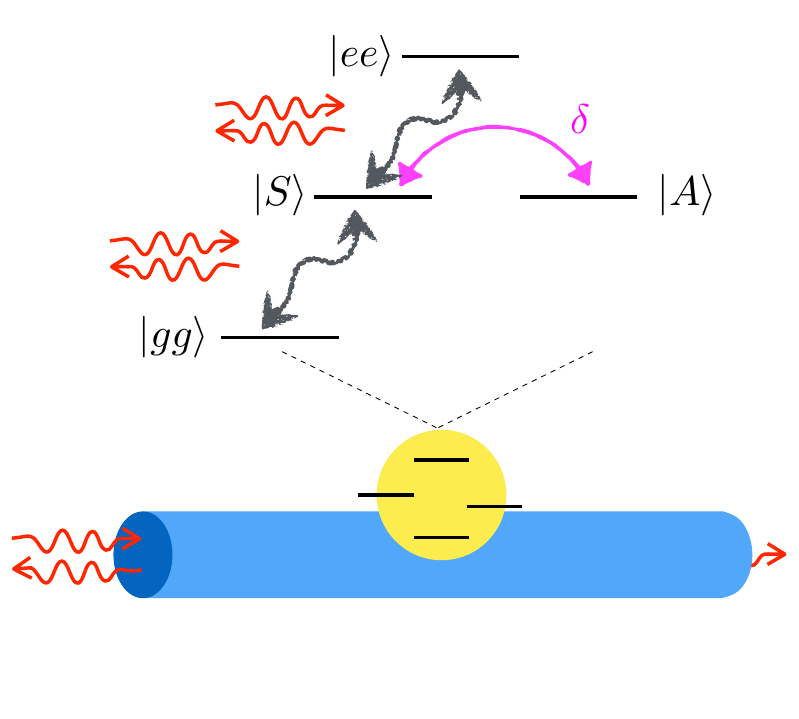}
	\caption{Schematic of the level structure of two detuned 2LS with $k_0L=2\,n\pi$ in the symmetric-antisymmetric basis. The effective $\Lambda$-type 3LS is formed from the three states $|gg\rangle$, $|S\rangle$, and $|A\rangle$. In the case of two incident photons at the resonant frequency $\omega_0$, the doubly excited state $|ee\rangle$ is not populated because of interference similar to that causing EIT. 
		\label{fig:mapping}}
\end{figure}

The mapping is made more transparent by rewriting $\mathcal{H}^0_\text{2LS}$ in the symmetric-antisymmetric (S-A) basis. Defining 
$\sigma_{S-}=(\sigma_{1-}\!+\!\sigma_{2-})/\sqrt{2}$ and 
$\sigma_{A-}=(\sigma_{1-}\!-\!\sigma_{2-})/\sqrt{2}$, one finds
$\mathcal{H}^0_\text{2LS} = \omega_0\sum_{i=S,A}\sigma_{i+}\sigma_{i-} 
+ \delta/2(\sigma_{S+}\sigma_{A-}+ \text{h.c.})$. 
Moreover, when $k_0L=n\pi$, by transforming to the momentum basis, we find that either the symmetric or the antisymmetric operator couples to the  photons (depending on whether $n$ is even or odd) but not both. Consequently, the operator structure for the pair in the S-A basis is precisely in the form of a driven $\Lambda$-3LS---see Fig.\,\ref{fig:mapping} for an illustration. However, an important difference remains: a pair of 2LS can hold up to two photons, $\sigma_{S+}\sigma_{S+}=-\sigma_{A+}\sigma_{A+}=\sigma_{1+}\sigma_{2+}\neq0$, while a driven 3LS can absorb only one photon at a time. As we discuss below, this difference impacts higher-order quantities.

\begin{figure}[t]
	\centering
	\includegraphics{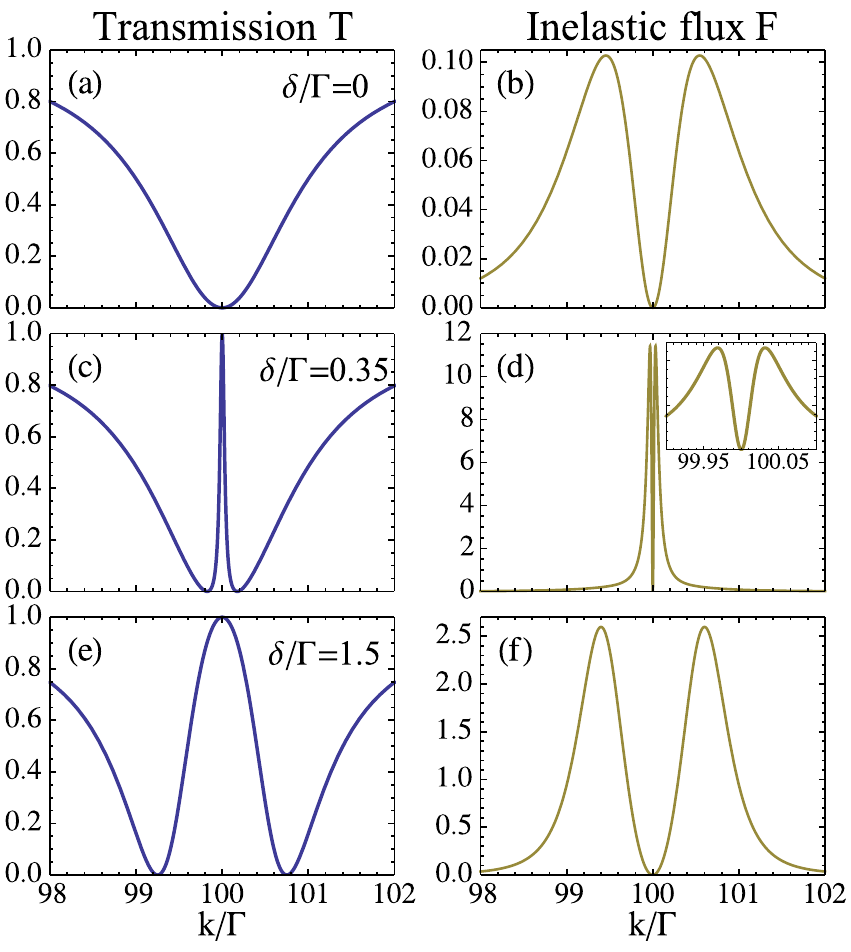}
	\caption{Pair of detuned 2LS: Single-photon transmission spectrum $T(k)$ (left column) and two-photon inelastic flux $F(k)$ (right column) as a function of incident momentum $k$ for a separation of $\lambda_0/2$ (so $k_0 L=\pi$). The detuning $\delta/\Gamma$ is set to 0 (top row), 0.35 (middle row, EIT regime), and 1.5 (bottom row). In the right column, red dotted curves represent transmitted flux $F_\text{R}$, and brown solid curves represent total flux $F=F_\text{R}+F_\text{L}$. Note that (i) $F=0$ at $k=\omega_0$, (ii) $T=0$ at $k=\omega_0\pm\delta/2$, and (iii) $T(\omega_0)=1$ when $\delta\neq0$. The inset in panel (d) magnifies the peaks around $\omega_0=100\Gamma$.
		\label{fig:detuned_2LS_T_F}}
\end{figure}

The single-photon transmission $T(k)\!=\!|t(k)|^2$ using Eq.\,\eqref{eq:transmission amplitude of a pair of 2LS} is shown in Fig.\,\ref{fig:detuned_2LS_T_F} for several detunings $\delta$. For zero detuning, one obtains a broad transmission minimum as expected based on the characteristics of a single 2LS. However, for non-zero detuning, a clear $T = 1$ EIT-like peak appears at the resonant frequency $\omega_0$, even though no 
external pumping is present. The transmission does still reach zero at certain frequencies---the $T\!=\!0$ dips are at $\omega_{1,2}=\omega_0\pm\delta/2$ since each individual 2LS is perfectly reflecting at its resonant frequency.

\subsection{Two-photon sector}
\label{subsec:2photon}

We now turn from the single-photon to the \emph{two-photon} sector. We find the response 
in this sector by computing the two-photon scattering wavefunction $|\psi_2(k)\rangle$, where $k$ is the wavevector of both incoming photons, and then extracting experimental observables  \cite{ZhengPRL13,FangEPJQT14,FangPRA15}. Two important observables are the two-photon correlation function (second-order coherence) in the transmission channel, 
\begin{equation}
g_2(t)\equiv \frac{\langle \psi_2| a_\text{R}^\dagger (x)a_\text{R}^\dagger(x+t)a_\text{R}(x+t)a_\text{R}(x)|\psi_2\rangle}{|T(k)|^2} ,
\end{equation}
and the inelastic power spectrum (resonance fluorescence), 
\begin{equation}
S_\alpha(\omega)\equiv\int dt e^{-i\omega t}\langle \psi_2 | a_\alpha^\dagger(x)a_\alpha(x+t)|\psi_2\rangle ,
\end{equation} 
where $\alpha=\text{R\;or\;L}$, with the elastic scattering 
delta-function removed.
The total inelastic scattering for incoming photons of momentum $k$, 
\begin{equation}
F=F_\text{R}+F_\text{L} \;\; \text{with}\;\;
   F_\alpha(k) \equiv \int d\omega\, S_\alpha(\omega) ,
   \quad
\end{equation}
is a valuable figure of merit for the strength of photon-photon interaction \cite{FangPE16} since energy exchange between the photons is a hallmark of such interaction. $F_\alpha(k)$ can then be used to compare different structures with the aim of maximizing interaction and correlation effects \cite{FangPE16}. 

It is known that due to interference effects, a 3LS has a fluorescence quench ($F=0$) at the EIT resonance $\omega_0$ regardless of the driving Rabi frequency $\Omega$ \cite{ZhouPRL96,FleischhauerRMP05,FangPE16,SanchezPRA16}. 
The lack of 
inelastic scattering implies that the ``bound state'' part of the wavefunction is absent \cite{ShenPRL07, ShiSunLSZ_PRB09, ZhengPRA10, RoyPRA11}. Therefore, at the EIT transmission resonance ($k=\omega_0$), there is no photon correlation and $g_2(t)=1$ for all times \cite{ZhengPRA12,RoyPRA14,FangPE16}.

For a pair of detuned 2LS, Fig.\,\ref{fig:detuned_2LS_T_F} shows that $F(k)$ behaves similarly. In principle, the system can absorb two photons and enter the doubly excited state $|ee\rangle$ (see Fig.\,\ref{fig:mapping}). However, the coherent interaction between the states $|S\rangle\equiv\sigma_{S+}|gg\rangle$ and $|A\rangle\equiv\sigma_{A+}|gg\rangle$, where $|gg\rangle$ is the ground state, blocks the occupation of $|ee\rangle$ in the steady state (for a time-dependent problem, there is transient occupation of $|ee\rangle$). To see this, consider the case in which $|S\rangle$ couples to the waveguide while $|A\rangle$ does not. In the steady state, because the system is $\Lambda$-3LS-like in the single photon sector, $|A\rangle$ is populated while the population of $|S\rangle$ is zero. A second photon cannot excite  
$|ee\rangle$ because $\sigma_{S+}\sigma_{A+}=0$. Alternatively, one can consider the different paths to reach $|ee\rangle$; then, the interference among them leads to zero occupation. Indeed, we find $\langle ee|\psi_2\rangle=0$ at $k=k_0$: \emph{in the steady state, the system is $\Lambda$-3LS-like in the two-photon sector as well.}  

In Figs.\,\figurepanel{fig:detuned_2LS_T_F}{d,f}, notice in this regard (i) the dip to $F\!=\!0$ on resonance; (ii) the similarity between $F$ in this few emitters, few photons scenario and the well-known absorption profile of a dense gas of $\Lambda$-3LS probed by laser beams \cite{FleischhauerRMP05}; and (iii) the symmetry of $F$ with respect to $\omega_0$. If the coupling of the qubits were not symmetric, the resonant frequency and the corresponding fluorescence quench would be offset from $\omega_0$ as given by Eq.\,\eqref{eq:mapping rules}.

\begin{figure}[t]
	\includegraphics[width=0.85\linewidth]{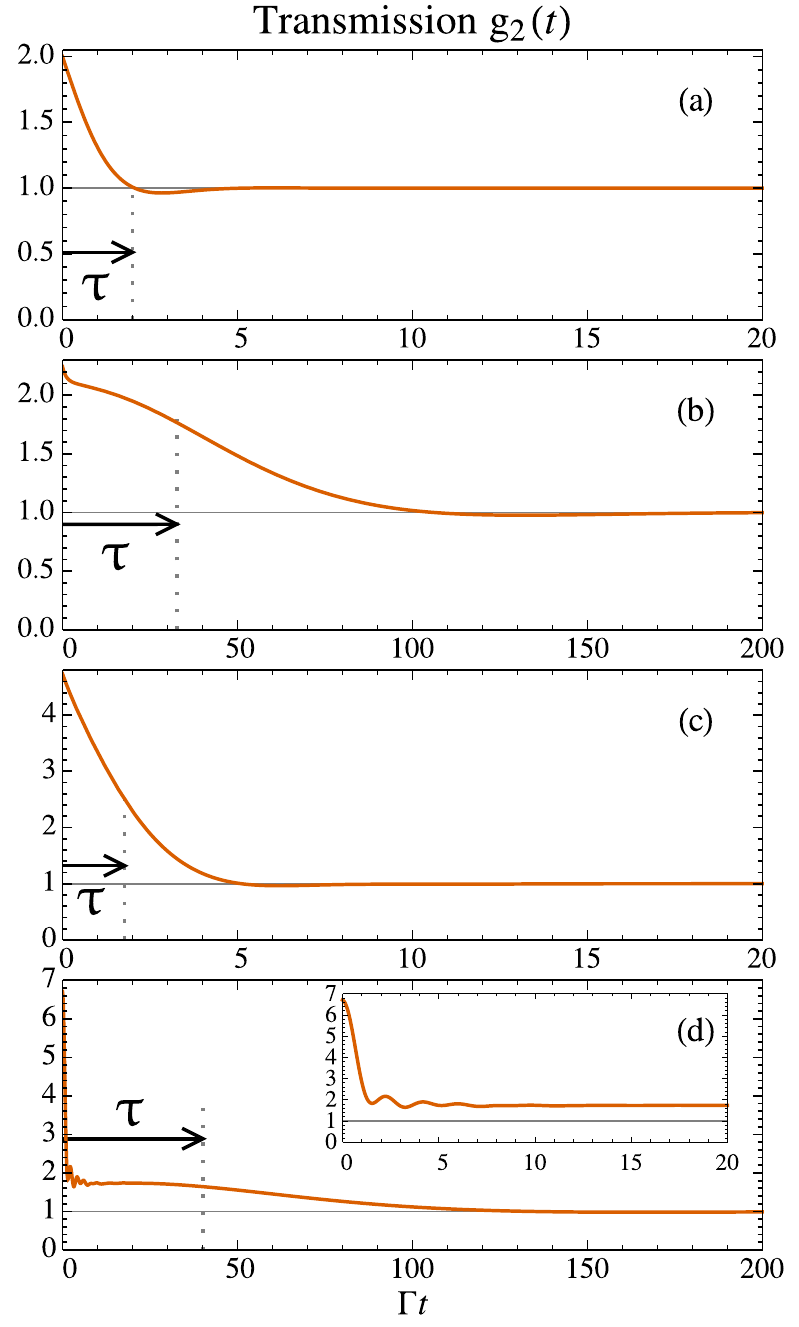}
	\caption{The two-photon correlation function $g_2(t)$ in the transmission channel for (a-c) a pair of detuned 2LS and (d) a 2-3-2 structure. The detunings are (a) $\delta/\Gamma=0$, (b) $\delta/\Gamma=0.35$, (c) $\delta/\Gamma=1.5$, and (d) $\delta/\Gamma=0.35$, in which $\Omega/\Gamma=3$ as well.  The incident frequency $k$ is chosen for each case so that $T=50\%$ and $k$ is on the red-detuned side of the $\omega_0=100\Gamma$ resonance. [The resonance itself, $k=\omega_0$ is not used because for case (a) $g_2(t)$ is ill-defined because in the elastic channel all photons are reflected, $T(k)=0$, while for cases (b) and (c), $g_2(t)=1$ (gray line).] The characteristic time delay $\uptau$ for each case is labeled. For panel (a), we use the fact that  $\uptau=2/\Gamma$ for an array of identical 2LS \cite{FangPRA15}. The separation between the two 2LS in all cases is $L=\lambda_0/2$.
		\label{fig:detuned_2LS_g2}}
\end{figure}

The correlation function $g_2(t)$ for two detuned 2LS is also similar to that of a driven $\Lambda$-3LS. Our results are shown in 
Fig.\,\ref{fig:detuned_2LS_g2}. When $\delta$ is small but not zero, the time delay $\uptau=d\arg[t(k)]/dk$ associated with the narrow resonance is large \cite{3LStime-delay}, so $g_2$ decays slowly. At the resonance $k=\omega_0$, because the scattering is entirely elastic ($F\!=\!0$), the photons cannot be correlated. Indeed, our calculation yields $g_2(t) \equiv 1$ identically for all time. 

The equivalence between a pair of detuned 2LS separated by $L=n\lambda_0/2$  and a driven $\Lambda$-3LS is, then, established in both the single- and two-photon sectors. (As this result depends upon the Markovian approximation, $\Gamma L/c \ll 1$, the value of $n$ is limited.)  Thus, using weak coherent states, one could experimentally obtain EIT-like properties in situations where driving is inconvenient.

\subsection{Identical 2LS} 
\label{sec: identical 2LS}

The special case $\delta=0$, Figs.\,\figurepanel{fig:detuned_2LS_T_F}{a,b} and \figurepanel{fig:detuned_2LS_g2}{a}, requires additional interpretation. Note that in this case one of the states $|A\rangle$ or $|S\rangle$ is completely decoupled, suggesting that there must be a change in properties. The mapping rules Eq.~\eqref{eq:mapping rules} imply that the zero-detuning case behaves like a single 2LS with the decay rate doubled. This is indeed true in the single-photon sector: in Fig.\,\figurepanel{fig:detuned_2LS_T_F}{a}, $T(k)$ has an inverse Lorentzian dip with width $2\Gamma$, agreeing with the fact that a 2LS behaves like a mirror at resonance \cite{RoyRMP17,LiaoPhyScr16}.

In contrast, this is \emph{not} true in the two-photon sector---a single 2LS \cite{LoudonQTL03} does not exhibit the fluorescence quench shown in Fig.\,\figurepanel{fig:detuned_2LS_T_F}{b}. This can be understood by first noting that the Markovian approximation causes the system to be periodic in $L$ such that the behavior for $k_0L=\pi$ is exactly the same as for $k_0L=0$, corresponding to two colocated 2LS. It is known that the fluorescence is quenched on resonance for two colocated, identical 2LS \cite{RephaeliPRA11}, thereby explaining the $F=0$ dip in Fig.\,\figurepanel{fig:detuned_2LS_T_F}{b}. 
Similarly, no correlation between the photons is generated on resonance, $g_2(t)=1$ for the reflected photons \cite{ShiPRA15} (since $T\!=\!0$ and $F\!=\!0$, everything is elastically reflected), a result that we reproduce (not shown). 
These effects result from interference \cite{ShiPRA15} involving the fourth level $|ee\rangle$ (though see discussion of the effect of loss below): there is an EIT-like three level interference in the two-photon sector. 
 
As a result, the mapping works only when $\delta\neq0$; when $\delta=0$, the two-photon behavior of a pair of 2LS and a $\Lambda$-3LS are different.

\section{Rectification Linked to Inelastic Scattering}
\label{sec: rectification}

\begin{figure*}[t]
	\centering
	\includegraphics[width=1.0\textwidth]{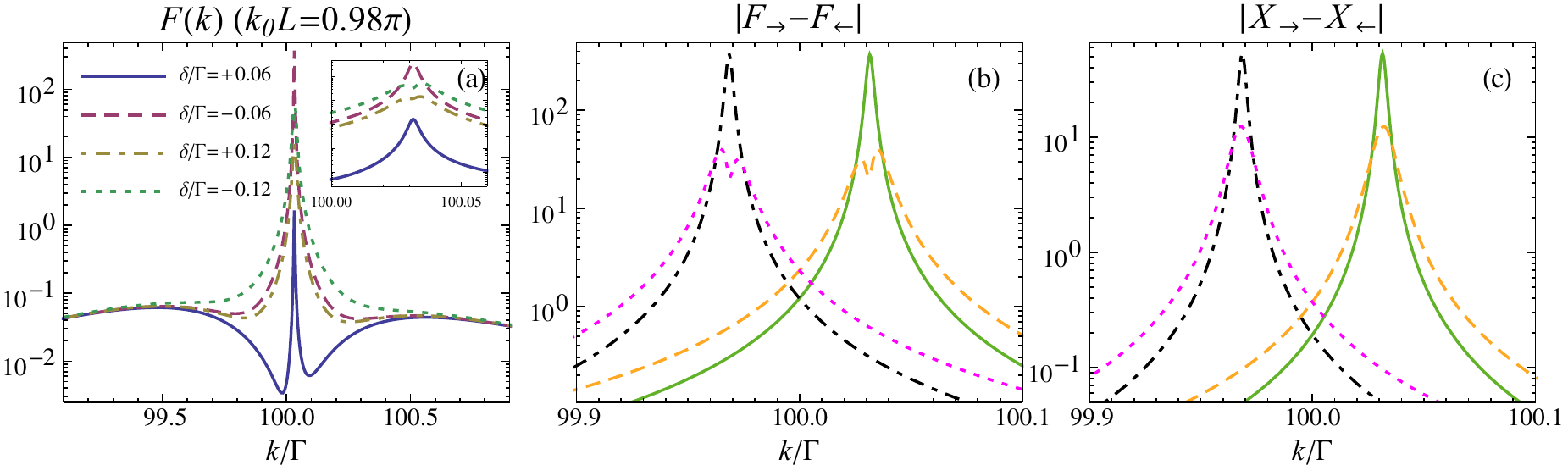}
	\caption{Non-reciprocal effects for a pair of 2LS as a function of incident frequency $k$. 
	(a) Transmitted inelastic flux $F(k)$ for several qubit detunings $\delta$. The peak position is at $\omega_0-\delta_\text{opt}/2\approx100.03\Gamma$. The inset magnifies the peaks. (b) Difference of transmitted flux due to left- ($\rightarrow$) and right-driving ($\leftarrow$) for two $L$ and two $\delta$. $(k_0L, \delta/\Gamma)=(0.98\pi, -0.06)$ (solid green), $(0.98\pi, -0.12)$ (dashed yellow), $(1.02\pi, +0.06)$ (dot-dashed black), and $(1.02\pi, +0.12)$ (dotted magenta). (c) Unnormalized rectification factor given by the difference in transmitted intensity. The color scheme and parameters are the same as in (b). $\omega_0/\Gamma=100$ in all panels.
		\label{fig:F_X_compare}}
\end{figure*}

We now turn to discussing rectification produced by two detuned 2LS. We have seen that non-reciprocity does not occur in the one-photon sector [Eq.\,(\ref{eq:transmission amplitude of a pair of 2LS})], and so we naturally turn to the two-photon sector, asking under what circumstances does substantial rectification occur.

We first note that in the colocated case ($L=0$) there can be no rectification for any number of photons---a point-like system cannot break left-right symmetry. Since for $k_0L=n\pi$ within the Markovian approximation, the properties of the system are the same as for the colocated case, for these spacings there likewise cannot be any rectification for any number of photons. Indeed, we check that our two-photon results are symmetric upon interchanging qubits 1 and 2 (see \cite{supplement} for the demonstration). This is consistent with calculations using other approaches \cite{DaiPRA15,FratiniPRA16,MascarenhasPRA16,disagreeFratini14} for these conditions, and reinforces the conclusion of Sec.~\ref{subsec:2photon} that two detuned 2LS with this separation behave like a driven $\Lambda$-3LS.  

If the separation between the two 2LS deviates slightly from  $k_0L=n\pi$, it is known, however, that non-reciprocal effects do occur \cite{DaiPRA15,FratiniPRA16,MascarenhasPRA16}. To study these cases, we begin by reiterating that our scattering theory approach is completely equivalent to input-output theory when a small coherent state amplitude $A$ is used \cite{FangPRA15}. For the purpose of comparing with previous work, therefore, we focus on results at the lowest power we are aware of, namely those in Fig.~2(d) of Ref.~\cite{DaiPRA15} \cite{dictionary}.
It can be seen that rectification is maximized roughly along a straight line in the $\delta$-$L$ plane, given through fitting by
\begin{equation}
	\delta/\Gamma \approx 3.18\Big(\frac{L}{\lambda_0/2} - 1\Big) .
\label{eq:fitted line}
\end{equation}
Therefore, we consider the situation slightly away from $L=\lambda_0/2$ (choosing $n=1$ here for simplicity).

Under the Markovian approximation, the transmission amplitude \eqref{eq:transmission amplitude of a pair of 2LS} has two poles:
\begin{equation}
\omega_\pm=\omega_0-\frac{i \Gamma }{2}\pm\frac{1}{2} \sqrt{\delta ^2-\Gamma ^2 e^{2 i k_0 L}}.
\end{equation}
In the EIT regime ($k_0L \approx \pi$ and $\delta/\Gamma\ll1$),
we can expand the poles $\omega_\pm$ to first order in $\delta$ and $L$, obtaining
\begin{equation}
\omega_\pm=
\begin{cases}
\omega_0-\frac{\Gamma}{2}(k_0L-\pi),& \quad \text{(dark)}\\
\omega_0+\frac{\Gamma}{2}(k_0L-\pi)-i\Gamma.& \quad \text{(bright)}
\end{cases}
\end{equation}
In addition, both poles have a $\mathcal{O}(\delta^2)$ contribution to the imaginary part (omitted) so that the dark pole is not fully decoupled and can be driven.
Recall that in the rectification setup usually considered \cite{FratiniPRL14} the incident frequency $k$ is on resonance with the second qubit, $k=\omega_0-\delta/2$. In order to drive the dark pole, $L$ and $\delta$ must then satisfy  
\begin{equation}
	\delta_\text{opt}/\Gamma \approx \pi\Big(\frac{L}{\lambda_0/2} - 1\Big) ,
\label{eq:rectification line}
\end{equation} 
which is very close to the fitted line (\ref{eq:fitted line}). 
The fact that the (nearly) dark pole has a very small imaginary part has two immediate implications. First, since the response time scale of the system is long, the system should be extraordinarily sensitive to the incident power, i.e.\ the number of photons arriving in that long time scale. 
Second, to get the maximum rectification one should tune very close to the dark pole 
\footnote{From this point of view, the necessity for ``extreme fine-tuning to select those working points'' noted in \cite{DaiPRA15} is natural.}. \emph{Therefore, rectification is maximized when the driving frequency matches the real part of the dark pole}.

This is an example demonstrating how the single-particle poles affect higher-order quantities: one-photon transport, corresponding to the linear response, is not sensitive to the slight shift of one-photon poles, but two- and multiple-photon transport are. Therefore, higher-order quantities such as $F(k)$ are suitable tools for understanding the non-reciprocity. 

Non-reciprocity is often quantified in terms of a rectification factor which is simply the difference in transmission from the left and right divided by their sum. From our two-photon scattering wavefunction, for a system driven from the left, the transmitted photon intensity for $x\gg 0$ is \cite{FangPRA15}
\begin{equation}
   _\text{RR}\langle \psi_2(k)| a_\text{R}^\dagger(x) a_\text{R}(x)|\psi_2(k)\rangle_\text{RR}
	= |t(k)|^2\frac{\delta(0)}{\pi}+X_\rightarrow(k) ,
\end{equation}
where the subscript ``$\rightarrow$'' denotes the driving direction and $X(k)$ is the sum of $F(k)$ and an interference term as described in Appendix \ref{appendix: X(k)}. It is clear that the first (second) term represents the linear (nonlinear) transport. Similarly, if the system is driven from the right, the transmitted intensity for $x\ll0$ is 
\begin{equation}
_\text{LL}\langle \psi_2(k)| a_\text{L}^\dagger(x) a_\text{L}(x)|\psi_2(k)\rangle_\text{LL}
= |t(k)|^2\frac{\delta(0)}{\pi}+X_\leftarrow(k) .
\end{equation}
The rectification factor $\mathcal{R}$ is therefore proportional to the difference in $X$,
\begin{equation}
\mathcal{R} \propto |X_\rightarrow(k)-X_\leftarrow(k)|,
\end{equation}
which can be calculated in our scattering theory approach. Note that we can only obtain an unnormalized rectification factor because the scattering states are delta-normalized and so the infinity $\delta(0)$, representing the ``volume'' of the system, appears in the expressions \cite{FangPRA15}.

In Fig.~\ref{fig:F_X_compare}, we present results for $F(k)$ and $|X_\rightarrow(k)-X_\leftarrow(k)|$ for parameters chosen both on and off the optimal straight line \eqref{eq:rectification line}. 
Fig.~\figurepanel{fig:F_X_compare}{a} shows clearly the substantial inelastic scattering for $\delta_\text{opt}$: $\delta_\text{opt}/\Gamma=-0.06$ for $k_0L=0.98\pi$. $F$ peaks at a very large value when the driving frequency matches the second qubit, $k=\omega_0-\delta_\text{opt}/2$ (see \cite{FangPE16} for values of $F$ in other systems). 

These conclusions carry over directly to the left-driving/right-driving difference in both $F$ and $X$ shown in Figs.~\figurepanel{fig:F_X_compare}{b,c}. These non-reciprocity measures do indeed peak at the detuning given by \eqref{eq:rectification line} when the energy of the incident photon matches the second qubit. For non-optimal $\delta$, there are peaks as a function of driving frequency but they are considerably smaller (note the logarithmic scale) than those at $\delta_\text{opt}$. Note that in these non-optimal cases, the peak is near the frequency of the dark pole not at the frequency of the second qubit---further indication that it is the dark pole that controls rectification. 

We note in passing that the region of large rectification is, strictly speaking, not an extended line on the $\delta$-$L$ plane, but rather a collection of segments. The reason is three-fold: (i) under the Markovian approximation, the system has periodicity $L$, so such a segment would appear around $L=n\lambda_0/2$; (ii) when $\delta$ is large, neither pole is  dark and so only small rectification is expected, following the trend in Fig.~\ref{fig:F_X_compare}; (iii) for large $L$, where the Markovian approximation no longer holds, an infinite number of poles will appear in Eq.~\eqref{eq:transmission amplitude of a pair of 2LS}, making the analysis difficult. This is an interesting regime which remains largely unexplored (see, however, \cite{LaaksoPRL14}). Because of the proliferation of poles, there is the possibility to achieve strong rectification at multiple frequencies. 

The main conclusions of this section are, then, that inelastic scattering is an inevitable ingredient for designing network components such as a photonic rectifier that rely on nonlinear transport of photons, that working at maximized $F(k)$ is often the optimal choice \cite{FangPE16}, and that rectification results from driving a dark pole.

\section{Recovering the fluorescence: 2-3-2} 
\label{sec: 2-3-2}

\begin{figure}[b]
	\centering
	\includegraphics{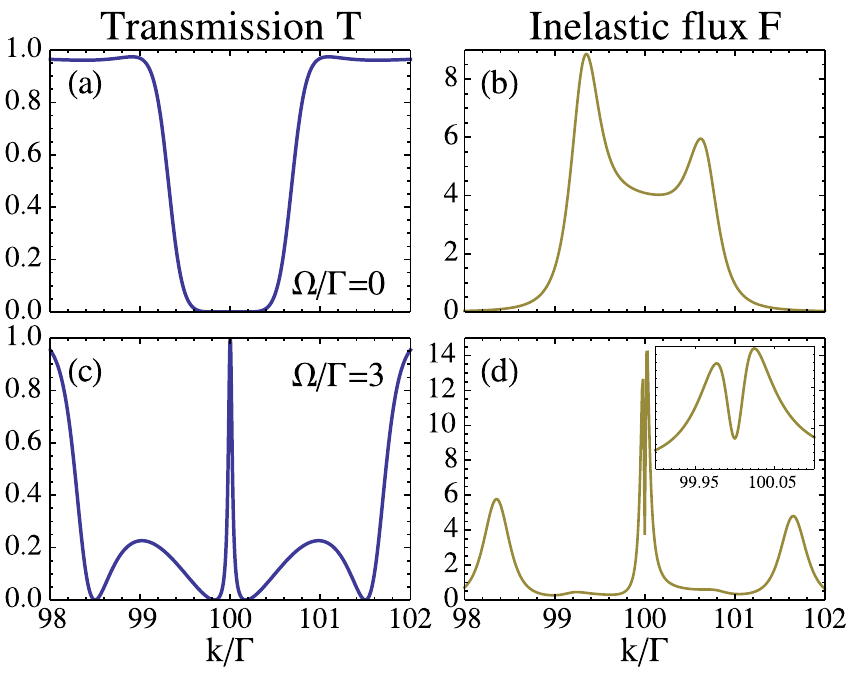}
	\caption{For the 2-3-2 Structure: Single-photon transmission spectrum $T(k)$ (left column) and two-photon inelastic flux $F(k)$ (right column) as a function of incident momentum $k$ for 2LS-3LS-2LS separated by $\lambda_0/4$ (so $k_0 L=\pi$). The detuning is $\delta/\Gamma=0.35$, and the Rabi frequency of the classical beam $\Omega/\Gamma$ is 0 (top row) and 3 (bottom row). In the right column, red dotted curves represent transmitted flux $F_\text{R}$, and brown solid curves represent total flux $F=F_\text{R}+F_\text{L}$. Note that (i) $F\neq0$ at $k=\omega_0$, (ii) $T=0$ at both $k=\omega_0\pm\delta/2$ and $\omega_0\pm\Omega/2$, and (iii) $T(\omega_0)=1$ when $\Omega\neq0$. The inset in panel (d) magnifies the peaks around $\omega_0=100\Gamma$.
		\label{fig:2-3-2_T_F}}
\end{figure}

\begin{figure}[t]
	\centering
	\includegraphics[width=0.85\linewidth]{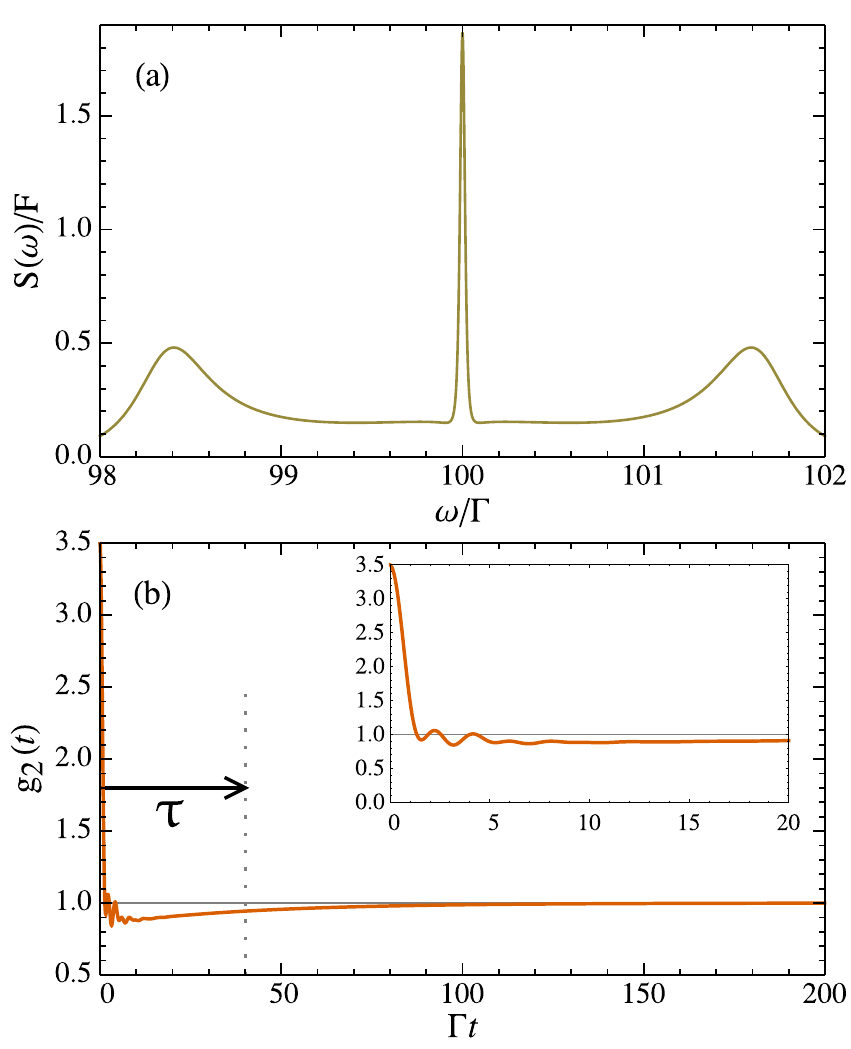}
	\caption{For the 2-3-2 structure excited at resonance, (a) the two-photon normalized power spectrum $S(\omega)/F$ as a function of output frequency $\omega$, and (b) the two-photon correlation function $g_2(t)$ in the transmission channel. In (a), both the transmitted fluorescence $S_\text{R}/F$ (red dotted line) and the total fluorescence $S_\text{R}/F+S_\text{L}/F$ (brown solid line) are shown. In panel (b), the uncorrelated value $g_2=1$ is labeled by a gray horizontal line, and the characteristic time delay $\uptau$ is also labeled (see main text). The inset magnifies $\Gamma t \in [0, 20]$.  Parameters used are  $k=\omega_0=100\Gamma$, $L=\lambda_0/2$, $\delta/\Gamma=0.35$, and $\Omega/\Gamma=3$ such that $T=100\%$.
		\label{fig:2-3-2_S_g2}}
\end{figure}

We now turn to discussing our second 
structure, namely the 2-3-2 structure shown in Fig.\,\ref{fig:schematic}. We show that by combining  in this way the two elements discussed above, one can achieve \emph{both} perfect elastic transmission \emph{and} strong inelastic scattering with photon correlations. 

The transmission $T$ and total inelastic flux $F$ for the 2-3-2 structure, calculated as described above, are shown in  Fig.\,\ref{fig:2-3-2_T_F}. The insertion of an undriven $\Lambda$-3LS ($\Omega\!=\!0$) midway between the two 2LS disrupts the interference effect causing the EIT-like transmission peak [panel (a)]. Thus in this case there is a broad transmission minimum, which is a precursor to a photonic bandgap \cite{FangPRA15}. On the other hand, the classical driving of the 3LS causes an EIT-like feature to appear in transmission, $T=1$ at $k=\omega_0$ [panel (c)], which is natural since the EIT peak of the driven $\Lambda$-3LS and the EIT-like interference peak of the two 2LS coincide. 

In sharp contrast to the previous cases, however, at the EIT-like resonance \emph{the fluorescence is not quenched}, $F\approx3.75\neq0$ in Fig.\,\figurepanel{fig:2-3-2_T_F}{d}. The corresponding power spectrum at resonance is plotted in Fig.\,\figurepanel{fig:2-3-2_S_g2}{a}: a sharp peak at $\omega_0$ and two side peaks around $\omega_0\pm\Omega/2$ are seen. The fact that the non-trivial ``bound state'' part of the wavefunction (or equivalently the two-photon irreducible T-matrix) is nonzero at resonance is key to causing these effects, a feature which is not present in either the isolated $\Lambda$-3LS or pair of detuned 2LS.

Likewise, the photon correlation function $g_2(t)$ is not identically one: it is shown for $k=\omega_0$ in Fig.\,\figurepanel{fig:2-3-2_S_g2}{b}. There is clear bunching at short times, $g_2(0)\approx3.47$, followed by a slow decay of (oscillating) anti-bunching. The time delay, 
\begin{equation}
\uptau=(4\Gamma/\delta^2)+(2\Gamma/\Omega^2)+\Gamma(4\Gamma/\delta^2)(2\Gamma/\Omega^2), 
\end{equation}
is indicated in 
Fig.\,\figurepanel{fig:2-3-2_S_g2}{b}, as in the double 2LS case. The last term signals the importance of interference: the time delay is not simply accumulated component by component. 

\begin{figure}[tb]
	\centering
	\includegraphics[width=1.0\linewidth]{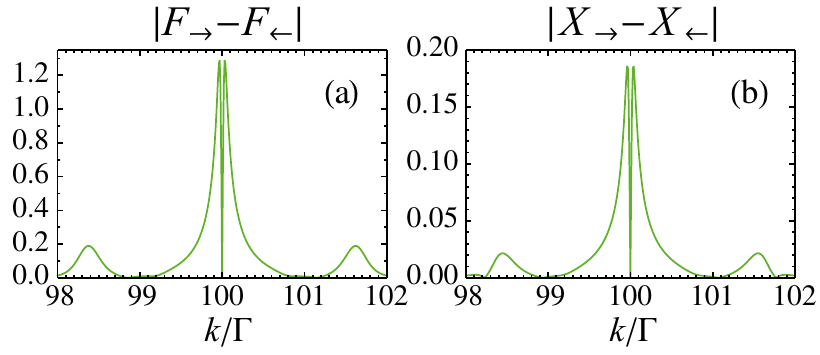}
	\caption{Non-reciprocal transmitted inelastic flux $F$ and intensity $X$ as a function of incident frequency $k$ for the 2-3-2 structure. (a) Difference of transmitted flux due to left- ($\rightarrow$) and right-driving ($\leftarrow$). (b) Unnormalized rectification factor given by the difference in transmitted intensity. Note that the value at $k=100\Gamma$ in both panels is zero. Parameters used are the same as those used in Fig.~\figurepanel{fig:2-3-2_T_F}{d}.
		\label{fig:2-3-2_rectification_example}}
\end{figure}

Several off-resonant features of the 2-3-2 results are interesting. First,  
$F$ is asymmetric about $k=\omega_0$ [Fig.\,\figurepanel{fig:2-3-2_T_F}{b,d}]: while in the previous cases $T$ and $F$ are both symmetric with respect to $\omega_0$ (see Fig.\,\ref{fig:detuned_2LS_T_F}), here only $T$ is symmetric. We find that interchanging the two 2LS causes 
$F(k)$ to reflect about $k=\omega_0$. The lack of left-right symmetry suggests non-reciprocity effects; indeed, Fig.~\ref{fig:2-3-2_rectification_example} shows that significant non-reciprocity and rectification occurs in this case for a range of input frequencies. The non-reciprocity effects have not been optimized here---we leave that for future study as there are many parameters for the 2-3-2 case. Nevertheless, as one point of comparison, if the separation between the emitters is doubled (separation between emitters of $\lambda_0/2$ so $k_0L=2\pi$), then $F(k)$ is symmetric about $k=\omega_0$, there is no inelastic scattering at that symmetry point, and there are no non-reciprocity effects.

The off-resonant photon correlations for the 2-3-2 are shown in Fig.\,\figurepanel{fig:detuned_2LS_g2}{d}, where $T=50\%$. $g_2(t)$ in this case has many similarities to that for a pair of detuned 2LS shown in panel (b). Note, however, several differences as well: oscillations are present due to interference between the three qubits, and the bunching at $t=0$ is substantially larger. 

These results suggest that structures that are slightly more complicated than the double 2LS, like the 2-3-2 structure considered here, may show enhanced rectification. Because there can be several dark poles in such more complicated structures even in the Markovian regime, this rectification may occur for a wider range of frequencies.

\section{Loss and Relaxing the Markovian Approximation}
\label{sec:loss-markovian}
 
We now briefly comment on two issues that have been neglected above. Both arise from the fact that the interference-induced transmission peaks here are ``composite'' in the sense that the transparency is due to scattering \emph{between} qubits, not \emph{within} the level structure of a single qubit as in EIT. This may in principle lead to different properties for certain structures, even if the nominal scattering characteristics are identical. In our double 2LS system, much as in the case of a lossy 2LS placed in a cavity \cite{WaksPRL06, ShenPRA09I}, the effect of decay to non-waveguide modes is quite different from that in EIT. 

\begin{figure}[tb]
	\centering
	\includegraphics[width=0.95\linewidth]{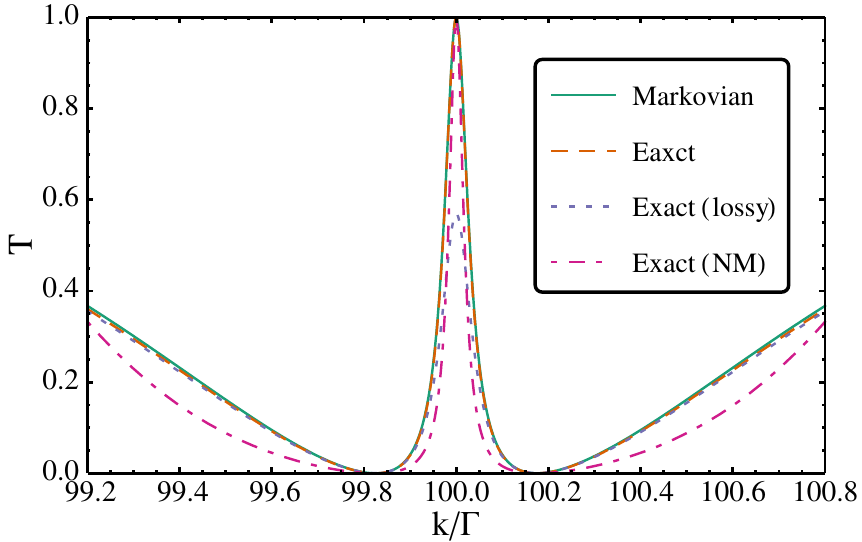}
	\caption{ 
		Assessing the lossless and Markovian approximations: the single-photon transmission $T(k)$ as a function of incident frequency $k$ for a pair of detuned 2LS with $\delta/\Gamma\!=\!0.35$. \emph{Loss:} the difference between the exact lossless (orange dashed, $L\!=\!\lambda_0/2$) and lossy (purple dotted) curves shows that modest loss has a significant effect on the EIT-like resonance---the peak transmission (at $k_0\!=\!100\Gamma$) decreases from $100\%$ to $T\approx56.9\%$ for $\Gamma'/\Gamma\!=\!0.02$.
\\\emph{Markovianity:} For $L\!=\!\lambda_0/2$ (and lossless), the Markovian result  (green solid) is nearly indistinguishable from the exact result. In contrast, for $L\!=\!20\lambda_0$, the exact result (magenta dot-dashed) is significantly different from the Markovian curve (which is the same for the two cases).     
		\label{fig:comparison_exact_markov}}
\end{figure}

In Fig.\,\ref{fig:comparison_exact_markov} we plot the transmission for a pair of 2LS separated by $\lambda_0/2$, with and without a loss rate of $\Gamma'$ to other channels (taken to be the same for both 2LS) \cite{loss-chiralchannel}. We use a large Purcell factor, $\Gamma/\Gamma' = 50$, since it is known that loss is small in superconducting circuits \cite{RoyRMP17}.
The transmission peak here is apparently not robust against loss, in sharp contrast to conventional setups where EIT is insensitive to loss in the excited state \cite{FleischhauerRMP05}.

This sensitivity to $\Gamma'$ can be understood using the S-A level structure discussed above (Sec.\ \ref{subsec:2photon} and Fig.\ \ref{fig:mapping}). In the lossless case, the asymmetric state $|A\rangle$ plays the role of the excited state with decay rate $2\Gamma$ (for $L\!=\!\lambda_0/2$), while the symmetric state $|S\rangle$ is metastable (zero decay rate). Loss $\Gamma'$ is added to both states when the mapping conditions Eq.~\eqref{eq:mapping rules} are satisfied. As a result, the ``metastable state'' is not more stable than the excited state with regard to the non-waveguide modes, causing the transmission peak to shrink. 

Finite loss affects not only the transmission peak but also the fluorescence quench and the correlations $g_2(t)$, as they are also caused by precise interference. It is known, for instance, that for identical colocated 2LS with loss ($\delta\!=\!0$, $L\!=\!0$, and $\Gamma'\neq 0$), inelastic scattering produces a ``Lorentzian-squared'' spectrum \cite{LalumierePRA13,lossLaLumiere} and causes correlation of the reflected photons, yielding $g_2(0)\approx 0$ instead of $1$ \cite{ZhengPRL13}. 

Finally, we briefly assess the importance of the Markovian approximation:  
Fig.\,\ref{fig:comparison_exact_markov} also shows the single-photon transmission in the lossless case calculated without using the Markovian approximation, Eq.~\eqref{eq:transmission amplitude of a pair of 2LS}. There is, as expected, no discernible difference for $L\!=\!\lambda_0/2$: the Markovian approximation works very well when the time for a round trip between the qubits is small enough compared to the inverse decay rates, $2L/c \ll 1/\Gamma$ \cite{LalumierePRA13,ZhengPRL13,LaaksoPRL14,FangPRA15}. 
However, when the qubits are separated by $L\!=\!20\lambda_0$, corresponding to 
$2L\Gamma/c \approx 2.5$, the exact curve and the Markovian approximation clearly differ.
Curiously, the effect of non-Markovianity is most clearly seen in the width of the resonant peak and the shape of the tails rather than the peak height.

There is increasing interest in quantifying and understanding general features of quantum non-Markovianity \cite{RivasRPP14,BreuerRMP16,deVegaRMP17}.
Recent developments in ``slow-light'' systems in which the group velocity is rather low, such as photonic crystal waveguides \cite{LauchtPRX12,GobanNatComm14} or surface acoustic waves \cite{GustafssonSci14,ArefBookChapter16}, suggest that waveguide QED provides a concrete arena in which non-Markovianity will play a role. Indeed, non-Markovianity measures have started being applied in waveguide QED \cite{TufarelliPRA14,RamosPRA16}, and it would be interesting to see further discussions in the present contexts. In this regard, we emphasize that our approach can handle the non-Markovian regime \cite{ZhengPRL13,FangPRA15,FangPE16} in the one- and two-photon sectors. 




\section{Conclusions}

The results presented here concern rectification, inelastic scattering, and photon correlations in two simple multiple-emitter quantum systems---a pair of detuned 2LS and the 2-3-2 structure. Our approach is to find scattering state solutions with either one or two incident photons, thereby connecting to weak coherent state excitation. 

We have shown that the response of a pair of 2LS separated by $L\!=\!n\lambda_0/2$ is identical to that of a driven $\Lambda$-3LS in both the single- and \emph{two-photon} sectors. In particular, there is an EIT-like transmission peak at which the fluorescence is quenched and the photons are uncorrelated. No rectification occurs. 

Tuning the separation slightly away from $n\lambda_0/2$, we showed that the ensuing rectification (which is solely in the two-photon sector) is caused by inelastic scattering associated with driving the dark pole of the two-2LS system. The location of this pole establishes a connection between the separation between the 2LS and detuning of their frequencies for which the non-reciprocity will be maximum. 

Finally, we showed that when a genuine driven $\Lambda$-3LS is inserted between the pair, inelastic scattering accompanied by photon correlation does occur. In this 2-3-2 system, then, one can create strongly correlated photons at perfect elastic transmission.

\begin{acknowledgments}
We thank Francesco Ciccarello, Io-Chun Hoi, Tao Shi, Xin Zhang, and Huaixiu Zheng for valuable discussions. This work was supported by U.S. NSF Grant No.~PHY-14-04125. 
\end{acknowledgments}

\appendix

\section{Transmission photon intensity}
\label{appendix: X(k)}

In this section we give explicit expressions for $X_\rightarrow(k)$ of the two-qubit system. Following the recipe of constructing the two-photon scattering wavefunction in Refs.~\cite{ZhengPRL13,FangEPJQT14,FangPRA15}, we have 
\begin{equation}
	X_\rightarrow (k) = \frac{F_\rightarrow(k)}{2\pi} + \text{(interference term)},
\end{equation}
where $F_\rightarrow(k)=\int d\omega S_\text{R}(\omega)$ is the transmitted inelastic flux for left-driving, and the interference term is
\begin{equation}
	-2\times\frac{e^{-ikx}}{\sqrt{4\pi}} t(k)^* \sum_{i,j} RR_i(k,x)(G^{-1})_{ij}e_j^\rightarrow(k)^2 + h.c.,
\end{equation}
where the matrix elements $G_{ij}$ are defined in Refs.~\cite{ZhengPRL13,FangEPJQT14}, 
$e_i^\rightarrow(k)$ is the wavefunction of qubit $i$ \cite{ZhengPRL13}: 
\begin{align}
	e_1^\rightarrow(k)&=\frac{\sqrt{\Gamma } e^{- i k L/2} \left[i (2k-2 \omega_0+\delta)-\Gamma  \left(1-e^{2 i k L}\right)\right]}{i\sqrt{\pi } \left[(2 k-2 \omega_0+i \Gamma)^2+\Gamma ^2 e^{2 i k L}-\delta ^2\right]},\\
	e_2^\rightarrow(k)&=\frac{\sqrt{\Gamma } e^{i k L/2} (2 k-2 \omega_0-\delta )}{\sqrt{\pi } \left[(2 k-2 \omega_0+i \Gamma)^2+\Gamma ^2 e^{2 i k L}-\delta ^2\right]},
\end{align}
and the function $RR_i(k,x)$ is defined as \cite{FangPRA15}
\begin{equation}
RR_i(k, x) = \frac{e_i^{\rightarrow}(k)^*}{\sqrt{\pi}}\int dq\,\frac{e^{iqx}\left[t(q)e_i^\rightarrow(q)^*+r^\leftarrow(q)e_i^\leftarrow(q)^*\right]}{k-q+i0^+}
\end{equation}
with $r(k)$ the reflection amplitude \cite{ZhengPRL13} 
\begin{equation}
	r^\rightarrow(k)=-i\Gamma\frac{4(k-\omega_0)\cos(kL)+2\left(\Gamma-i\delta\right)\sin(kL)}{(2 k-2 \omega_0+i \Gamma)^2+\Gamma ^2 e^{2 i k L}-\delta ^2}.
\end{equation}
The calculation of $X_\leftarrow(k)$ follows in a similar way. 


We close by three remarks: 
(i) the reflection amplitude $r(k)$ is not invariant upon exchanging qubit 1 and 2, as which qubit is hit first matters;
(ii) the amplitudes $r^\leftarrow(k)$ and $\{e_i^\leftarrow(k)\}$ for right-driving can be simply obtained by taking $r^\leftarrow(k)=r^\rightarrow(k)|_{\delta\rightarrow-\delta}$, $e_1^\leftarrow(k)=e_2^\rightarrow(k)|_{\delta\rightarrow-\delta}$ and $e_2^\leftarrow(k)=e_1^\rightarrow(k)|_{\delta\rightarrow-\delta}$, which exploits the mirror symmetry; (iii) all contour integrals, after the Markovian approximation is made, can be solved symbolically \cite{supplement} using Mathematica. However, we are unable to write down a concise closed form for the results for arbitrary $\delta$ and $L$; a limiting case ($\delta=0$) is given in Ref.~\cite{FangEPJQT14}.

\bibliography{WaveguideQED_bib/WQED_2016,BibFootnotes}

\begin{thebibliography}{75}
\bibitem [{3LS()}]{3LStime-delay}%
  \BibitemOpen
  \href@noop {} {}\bibinfo {note} {For a driven $\Lambda$-3LS,
  $\uptau=2\Gamma_e/\Omega^2$ at $k=k_0$. Thus, using the mapping rules
  Eq.~\eqref{eq:mapping rules}, one finds that $\uptau$ is $4\Gamma/\delta^2$
  for the case of a 2LS pair.}\BibitemShut {Stop}%
\bibitem [{dis()}]{disagreeFratini14}%
  \BibitemOpen
  \href@noop {} {}\bibinfo {note} {In contrast, note the disagreement with
  Ref.~\cite{FratiniPRL14}: our calculation implies that the rectifying factor
  $\mathcal{R}$ should be zero along the line cut $L=1$ in Fig.\,3(a) of
  \cite{FratiniPRL14}.}\BibitemShut {Stop}%
\bibitem [{dic()}]{dictionary}%
  \BibitemOpen
  \href@noop {} {}\bibinfo {note} {Dictionary: to compare with
  Ref.~\cite{DaiPRA15}, the translation to our notation is
  $\gamma\rightarrow\Gamma/2$, $\theta\rightarrow 2k_0L$,
  $\Delta\rightarrow-\delta$, and $\alpha\rightarrow A$. As there is a
  noticeable difference between the results in panels 2(d) and 2(c) in
  Ref.~\cite{DaiPRA15}, the power used for 2(c) is already too high for
  comparison to our two photon results.}\BibitemShut {Stop}%
\bibitem [{los({\natexlab{a}})}]{loss-chiralchannel}%
  \BibitemOpen
  \href@noop {} {} \bibinfo {note} {Formally, every lossy
  level coupled to the 1D continuum requires an independent fictitious chiral
  channel modeling the dissipative environment, but in the single-photon sector
  it is equivalent to a simple replacement
  $\omega\rightarrow\omega-i\Gamma'/2$.}\BibitemShut {Stop}%
\end{thebibliography}

\end{document}